\documentclass[aps,pre,twocolumn,showpacs,superscriptaddress,groupedaddress,10pt]{revtex4-1}  

\usepackage{graphicx}  
\usepackage{grffile}
\usepackage{dcolumn}   
\usepackage{bm}        
\usepackage{amssymb}   
\usepackage{amsmath}
\usepackage{amsthm}
\usepackage{amsthm}
\usepackage{mathtools}
\usepackage{xcolor}
\usepackage{tikz}	
\usepackage[ansinew]{inputenc}
\usepackage{soul}

\begin{document}



\title{Ferromagnetic Potts models with multi-site interaction}

\date{\today}

\author{Nir Schreiber}
\affiliation{Department of Mathematics, Bar Ilan University, Ramat Gan, Israel 5290002}
\author{Reuven Cohen}
\affiliation{Department of Mathematics, Bar Ilan University, Ramat Gan, Israel 5290002}
\author{Simi Haber}
\affiliation{Department of Mathematics, Bar Ilan University, Ramat Gan, Israel 5290002}
\begin{abstract}
We study the $q$ states Potts model with four site interaction on the square lattice.
Based on the asymptotic behaviour of lattice animals, it is argued that
when $q\leq 4$ the system exhibits a second-order phase transition,
and when $q > 4$ the transition is first order. The $q=4$ model is borderline.
We find ${1}/{\ln q}$ to be an upper bound on $T_c$, the exact critical temperature.
Using a low-temperature expansion, we show that $1/(\theta\ln q)$, where $\theta>1$ is 
a $q$-dependent geometrical term, is an improved upper bound on $T_c$.
In fact, our findings support $T_c=1/(\theta\ln q)$. This expression is used to
estimate the finite correlation length in first-order transition systems.
These results can be extended to other lattices.
Our theoretical predictions are confirmed numerically by an extensive study
of the four-site interaction model using the Wang-Landau entropic sampling method for $q=3,4,5$.
In particular, the $q=4$ model shows an ambiguous finite-size pseudocritical behaviour.
\end{abstract}
\maketitle
\section{Introduction}

The Potts model \cite{Potts1952,Wu1982} has been widely explored in the literature for the last few decades. While many analytical and numerical results exist for the traditional two-site interaction model in various geometries and dimensions \cite{Wu1982}, little is yet known about models with multisite interactions \cite{Baxter1978,Enting1982,Baxter1978,Wu1979,Wu1980,Wu1981}.
Baxter $et~al.$ \cite{Baxter1978} and Wu $et~al.$
\cite{Wu1979,Wu1980,Wu1981}
obtained the exact transition
point for the three-site interaction model on the triangular lattice. The four-spin interaction model has been studied by several authors \cite{Giri1977,Kunz1978,Burkhardt1979}. Specifically, it has been shown \cite{Giri1977,Kunz1978} that the site percolation problem
on the square lattice can be formulated as a four site interaction Potts  model in the limit $q\to 1$.
Burkhardt \cite{Burkhardt1979} argued that the four-site Hamiltonian $\cal H$, with
interaction strength $K$ defined
for every other square of the lattice (chequerboard), can be mapped onto
another four-site Hamiltonian $\tilde{\cal H}$ with strength $\tilde K$,
defined for every elementary square in the dual lattice. This mapping yielded the transformation
\begin{equation}
\label{eq:Essam_sq}
(e^{K}-1)(e^{\tilde K}-1)= q^3,
\end{equation}
in agreement with a more general expression \cite{Essam1979,Wu1982}
\begin{equation}
\label{eq:Essam_gen}
(e^{K_\gamma}-1)(e^{\tilde K_\gamma}-1)= q^{\gamma-1},
\end{equation}
which assumes arbitrary $\gamma$ site interaction.
Results like\ (\ref{eq:Essam_sq}) and\ (\ref{eq:Essam_gen}) may be conveniently
obtained if one equivalently represents the Potts spin configurations as
graphs on regular lattices \cite{Wu1982,Baxter1973a,Baxter1973b}.
However, the set of monochromatic
graphs associated with non-zero interaction terms in the checkerboard Hamiltonian,
is small compared to the set of monochromatic graphs
involved in the partition sum of a problem where every elementary square is
considered.
Therefore,\ (\ref{eq:Essam_sq}) suggests that
the transition
point (if it exists) should be rather different from that of a four site interaction model
defined for every
elementary square.

In this paper we consider a four site interaction model
described by a Hamiltonian with a partition sum that exhausts
all the elementary squares of the lattice.
We propose a simple equilibrium argument
that results in a critical condition for the transition point.
This condition
is in fact a zeroth-order approximation to the exact point.
It relies on the observation that tracing out spin states in the partition sum
is equivalent to the enumeration of large scale lattice animals at the
vicinity of the transition point.
Using a self consistent low temperature approximation, we 
obtain a more general condition which may allow one to approach the exact point up to an arbitrarily small distance by means of the first-order finite correlation length, {\it at least} when $q>4$.
It is argued that these considerations can be applied to other lattices. To demonstrate the generalization, we briefly also discuss the triangular lattice.
We next test our analytical predictions by an extensive numerical study of
 the four-site interaction Potts model on the square lattice (FPS) with $q=3,4,5$ states per spin.
For that purpose we use the Wang-Landau (WL) \cite{Wang2001,Wang2001a} entropic sampling method. The simulations results, together with finite size scaling (FSS) analysis, enable us
to approximate the infinite lattice transition point for each of the three models.
An estimate of the correlation length for the $q=5$ model, which according to the simulations exhibits a strong first-order transition,
is additionally made. 
It should be noted that another microcanonical-ensemble-based approach that may
be useful in simulating the first-order transition FPS has been introduced in \cite{MartinMayor2007}.

The rest of the paper is organized as follows. In Sec.\ \ref{sec:AR}
 we present the model and describe the role of lattice animals in determining the order of the phase transition. We find the (seemingly) exact transition point and show it is related to the finite
 correlation length in the first-order transition case. In Sec.\ \ref{sec:Sim} we
 present the WL simulations results and FSS analysis. Our conclusions are drawn in Sec.\ \ref{sec:con}.
  	
\section{Analytical results}
\label{sec:AR}
\begin{figure}
\centering
\vspace*{1cm}
\includegraphics[width=0.43\textwidth,natwidth=440,natheight=240]{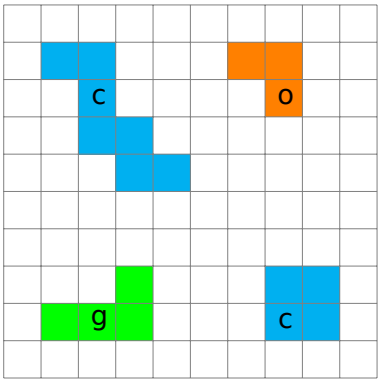}\hspace*{-2.5cm}
\caption{\label{fig:animals}A portion of the square lattice showing a graph $G$ 
with $c(G)=4$ monochromatic clusters, $f(G) = 18$ faces (coloured squares),
and $\nu(G) = 43$ nodes residing in the corners of these squares. The three different colours
(also denoted by c,g,o) represent a model with $q\geq 3$.}
\end{figure}

We consider the FPS, defined by the Hamiltonian
\begin{equation}
\label{eq:H}
-\beta {\mathcal H} = K\sum_\square\delta_{\sigma_\square},
\end{equation}
where $\beta=1/k_BT$ and $K=\beta J$ is the dimensionless coupling strength (for convenience we will assume from now on $k_B=J=1$). Each spin can take an integer value $1,2,...,q$. The $\delta_{\sigma_\square}$ symbol assigns $1$ if
all the four spins in a unit cell $\square$ are equal and $0$ otherwise.
The summation is taken over all the unit cells.
It is convenient to write
the partition function for
the Hamiltonian (\ref{eq:H}) \citep{0486462714}
\begin{eqnarray}
\label{eq:Z_N}
Z_N = \sum_{\sigma_\square}\prod_\square(1+v\delta_{\sigma_\square})
\sim q^N\sum_{G} q^{c(G)-\nu(G)}v^{f(G)},
\end{eqnarray}
where $v=e^K-1$ and $G$ is a graph made of $f(G)$ unit cell faces placed on the
edges of the lattice.
The faces are grouped into $c(G)$ clusters with a total number of $\nu(G)$ nodes.
The $\sim$ sign is due to  contributions to the partition sum from perimeter terms $o(N)$, which are omitted.
Clusters with perimeters of size $O(N)$ (snakelike, snail-like, etc.) are
energetically unfavourable and also assumed to be poor in entropy; therefore
their corresponding graph contributions are absent. 
An illustration of a graph $G$ is given in Fig.\ \ref{fig:animals}. Provided all the interacting spins are shown in the figure,
$G$ is associated with a $q^{N-39}v^{18}$ term in\ (\ref{eq:Z_N}). 

We now consider a low temperature expansion ($v\approx u=e^K$) in which we
assume only $k$ clusters exist. That is, for each $k$ large enough we
assume a single cluster [$c(G)=1$] with $f(G)=k$ faces and $\nu(G)=m_k$ sites. It is conjectured that in a typical $k$ cluster $m_k\approx k$. 
In terms of the new variables, the low temperature partition function may take the form
\begin{equation}
\label{eq:Z_N_large}
Z_N^{\rm low} \propto q^N\sum_k\sum_{m_k}{\cal G}(k,m_k)q^{-m_k}u^k,
\end{equation}
where ${\cal G}(k,m_k)$ is the number of configurations with $k$ faces and $m_k$ sites,
associated with a $k$ cluster. It is known \cite{Barequet2016,Klarner1973,Jensen2003}
that the combinatorial term $g_k = \sum_{m_k}{\cal G}(k,m_k)$ for large $k$
 is the asymptotic number of lattice animals 
 \footnote{ It can be shown that $k$ clusters and  clusters of size $o(N)$ do not share mutual 
 $\text{"corner"}$ sites, asymptotically almost surely.}
$g_k\approx c\lambda^k/k$, where  $\lambda\approx 4.0626$ and $c \approx 0.3169$. This observation distinguishes between
$q>4$ and $q \leq 4$.
Making a $k$ cluster (animal) monochromatic,
the total change in entropy if an asymptotic number of site configurations is exhausted, 
can be written, to leading order, as
\begin{equation}
\label{eq:Delta_S}
\Delta S_{\text{tot}}= k\ln(\lambda/q).
\end{equation}
Thus, when $q>\lambda>4$, it is energetically disadvantageous for the system to occupy animals at the asymptotic rate. Instead, to optimize the energy gain to entropy loss ratio,
it possesses a giant component (GC),
typically at the system size, that may be distorted from a perfect square in shape.
This mechanism is usually associated with systems which exhibit a first-order
phase transition.
In the case that $q\leq 4$,
since $\lambda>q$, the entropy of the system increases.
To avoid this, the system will again form a GC but this time with a fractal dimension rather than a
simple component as in the $q>4$ case. This scenario is typical to second-order
transitions, where the correlation length at criticality diverges.
A single monochromatic GC approximately reduces the entropy in the amount of $\Delta S= -\ln q^{k_{GC}+h.o.t}\approx -\ln q^{k_{GC}}$. The resulting gain in energy is $\Delta E=-k_{GC}$. Thus, $\Delta F=\Delta E-T\Delta S<0$ if and only if
$T<1/\ln q$, yielding the zeroth-order bound on the critical point
\begin{equation}
\label{eq:Tc}
\tilde T_c = \frac{1}{\ln q}.
\end{equation}

Consider for a first-order $q$ the class (denoted by $\hat A$) of large $k$ animals
with perimeters proportional (to leading order) to $\sqrt N$. 
Higher order contributions to\ (\ref{eq:Delta_S}) from 
the simple GC may then be depicted by writing
\begin{equation}
\label{eq:sup0}
\theta = \sup_k\left(\sup_{m_k}\frac{m_k}{k}\right),
\end{equation}
where $m_k$ are now site variables of animals in $\hat A$. Replacing $q^{-m_k}$ in (\ref{eq:Z_N_large}) with $q^{-\theta k}$, it can be shown (see Appendix\ \ref{app:Tc}) that
\begin{equation}
\label{eq:Lambda}
\Lambda = \lim_{N\to\infty}(Z_N^{\rm low})^{1/N}= u q^{1-\theta}.
\end{equation}
The (minus) dimensionless free energy
$-\beta f^{\rm low} =\ln\Lambda$ is then maximal if and only if $uq^{-\theta}>1$, leading to the critical condition
\begin{equation}
\label{eq:u_c0}
u_c=q^\theta,
\end{equation}
or equivalently to the critical temperature
\begin{equation}
\label{eq:Tc_mod}
\hat T_c =\frac{1}{\theta\ln q}.
\end{equation}
Note that if one does not adopt the low temperature approximation,
one has to add the term $\ln(1-1/u)$ to $-\beta f^{\rm low}$,
hence does not violate the critical condition (\ref{eq:u_c0}).
Note also that long range order is uniquely controlled by large animals.
These two observations imply that the critical temperature\ (\ref{eq:Tc_mod}) is exact.
Observe also the approximation $m_k=k$ (in
the exponent) in\ (\ref{eq:Z_N_large}) results in the critical condition
$u_c = q$ and likewise\ (\ref{eq:Tc}).
Equation\ (\ref{eq:Tc_mod}) can be used to relate the critical point to the
finite correlation length through
\begin{equation}
\label{eq:xi}
\theta = 1+c_1/\xi+...,
\end{equation}
where $\xi$ is a typical length for
clusters that are not $k$ clusters. For instance, for the square lattice, it can be easily shown that the simple GC consists of $k$ faces and $m_k$ sites
satisfying
\begin{equation}
\label{eq:ratio}
\frac{m_k}{k}\leq 1+\frac{\hat c}{\sqrt k}+...
\end{equation}
with $\hat c\geq 2$ constant. It follows from\ (\ref{eq:xi}),(\ref{eq:ratio}) (see Appendix\ \ref{app:xi}) that $c_1=\hat c$.
With the further aid of\ (\ref{eq:Tc_mod}), one readily obtains
\begin{equation}
\label{eq:Tc_xi}
\hat T_c(q,\xi) = \frac{1}{\ln q}\left(1-\frac{\hat c}{\xi}\right)+O(1/\xi^2).
\end{equation}

Finally, we address the issue of the lattice structure. In agreement with ref. \cite{Delfino2017},
the formation mechanism of a GC, either simple or fractal, which controls the critical properties of the model, applies also to other systems.
Specifically, the zeroth-order approximation (\ref{eq:Tc})
is expected to be valid (up to a constant multiplicative factor) for other lattices. In the first- order transition case, the lattice structure is captured by means of the constant term in\  (\ref{eq:Tc_xi}).
For example, in the triangular lattice, a simple GC consisting of $m_k = k/2 + O(\sqrt k)$ sites,
satisfies\ (\ref{eq:Tc_xi}) with $\hat c \geq 1$. 
The lower bound corresponds to the marginal case where the GC, when embedded in 
the square lattice, forms a perfect monochromatic square with no vacancies.

\section{Simulations}
\label{sec:Sim}
\begin{figure}
\includegraphics[width = \columnwidth,height = 7cm]{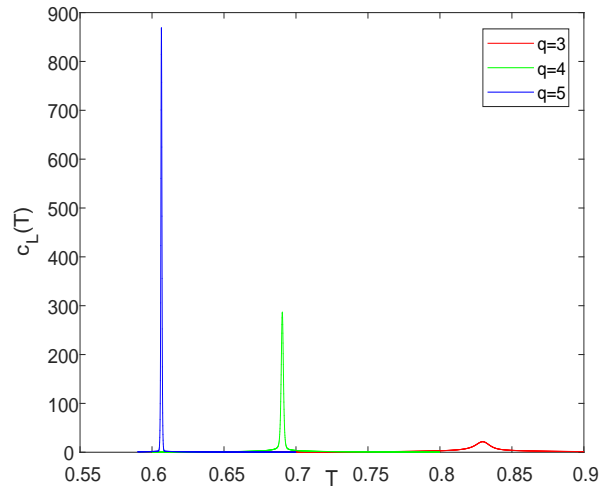}
\caption{\label{fig:Cv}
Variation of the specific heat of each model against
temperature for $L = 44$. While the $q = 4$ model and especially the
$q = 5$ model display sharp and narrow peaks at the $q$-dependent
position of the specific-heat maximum $T_L(q)$, the $q = 3$ peak is an
order of magnitude smaller and rather broad.}
\end{figure}

\begin{figure}[ht]
\includegraphics[width = \columnwidth,height = 7.5cm]{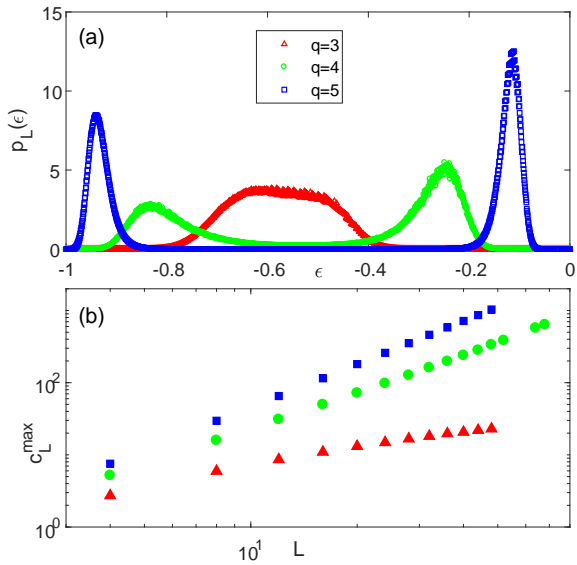}
\caption{\label{fig:energyPDF_3models}
(a) Pseudo-critical canonical energy distribution computed at
$T_L(q)$ for $q=3,4,5$ and $L=44$. Note the
peaks width $1/L$ behaviour when $q=5$, typical to normal distributions.
Conversely, the distributions for the $q=4$
(and of course the $q=3$) models are essentially not normal.
(b) Scaling of the specific heat maximum $c_L^{\max}(q)$ with $L$
on a log-log scale for $q=3$ (${\color{red}\blacktriangle}$), $q=4$ (${\color{green}\bullet}$)
 and $q=5$ (${\color{blue}\blacksquare}$).
}
\end{figure}

To test our analytical predictions, we study the FPS for three different models, namely, with
$q=3,4$ and $q=5$ states per spin.
The Wang-Landau (WL) \cite{Wang2001,Wang2001a} entropic sampling method is chosen for this purpose
since it enables one to accurately compute canonical averages at any desired temperature. We use lattices with linear size $L=4,8,12,...,68$ and periodic boundary conditions are imposed.
For each lattice size, we compute $\Omega(E)$, the number of states with energy $E$. These quantities allow us to
calculate energy-dependent moments $\langle E^n\rangle \propto \sum_E E^n \Omega(E)e^{-\beta E}$. In particular, we are interested in the specific heat per spin given by \cite{Newman1999,Stanley1987}
\begin{equation}
c_L = L^{-d}\beta^2(\langle E^2\rangle-\langle E\rangle^2).
\end{equation}
A plot of the specific heat for the three models is given in Fig.\ \ref{fig:Cv}.
For each model, the location of the peak serves as $L$-dependent pseudo-critical temperature and
is defined as $T_L\equiv T_{C_L^{\rm max}}$. Indeed, in agreement with\ (\ref{eq:Tc_mod}),
the pseudo-critical temperatures increase with $q$.
To determine the order of the transition for each model we
are simultaneously also interested in the energy probability density.
The latter may be written 
\begin{equation}
\label{eq:DOS}
P_L(\epsilon) \propto g_L(\epsilon)e^{-\beta L^d \epsilon}\approx
L^d \Omega(E)e^{-\beta E},
\end{equation}
with $\epsilon = L^{-d}E$ and $g_L(\epsilon)$ is the energy density of states. In Fig.\  \ref{fig:energyPDF_3models}a we display the probability density
at $T_L(q)$.
The $q=3,4$ models apparently suffer from significant finite-size effects.
Specifically, the $q=4$ model has a double-peaked shape, usually seen in first-order
transitions \cite{Janke1993}.
Evidently, there is a large dip between the peaks, but (unlike in the $q=5$ case) also a domain where the two humps overlap.
A fit of the minimal density between the peaks to a power law, generates
a slope $-1.09\pm 0.19$. 
This may indicate finite-size
interface contributions to the PDF. Either way, the dip does not exponentially vanish
as expected from systems which undergo a discontinuous transition.
When $q=5$, the energy is narrowly distributed in the
vicinity of the ordered and disordered states' energies
(denoted by $\epsilon_-$ and $\epsilon_+$ respectively), and has a typical
width $1/L$.

\begin{figure}
\includegraphics[width = \columnwidth,height = 8.6cm]{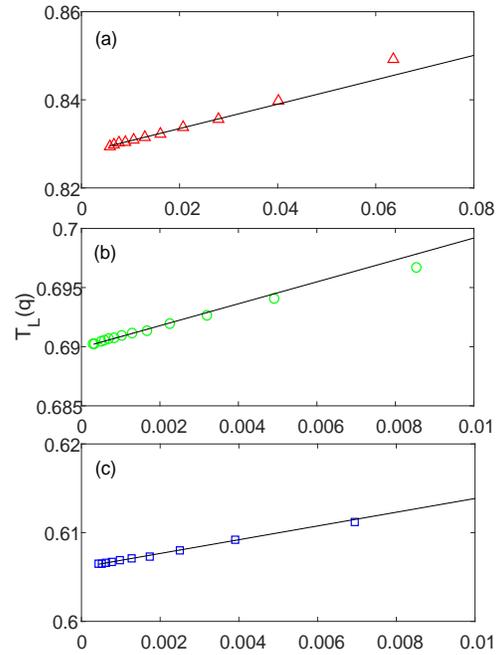}
\caption{\label{fig:Tc}
Scaling of the position [temperature $T_L(q)$] of the specific heat maximum with $L$,
 for the three models:
(a) $q=3$ and $T_L-T_c\propto L^{-1/\nu}(\ln L)^{\alpha/\nu}$. (b) $q=4$ and
$T_L-T_c\propto L^{-1/\nu+\text{h.o.t}}$. (c) $q=5$ and $T_L-T_c\propto L^{-d}$.
Solid lines are presented to guide the eye.}
\end{figure}

\begin{figure}[h]
\includegraphics[width = \columnwidth,height = 7cm]{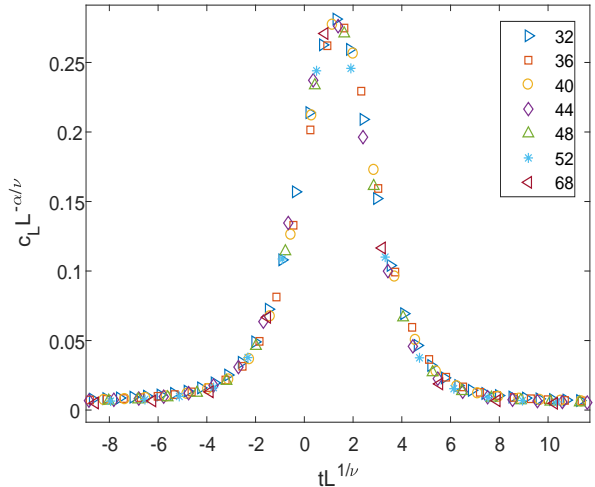}
\caption{\label{fig:CvUniversal}	
The specific heat universal scaling function ${\cal F}(x)$
for several lattice sizes $L$. The estimated values $T_c(4)\approx 0.689(9)$
and $\alpha/\nu\approx 1.832(7)$ are used in all the plots.}
\end{figure}

Armed with these= observations we next perform a FSS
analysis to each of the models.
For each $q$ we locate $c_L^{\max}(q)$ and $T_L(q)$. We fit these observables to linear models according to conventional scaling laws. We then vary $L_{\min}$, the smallest lattice size used in the fit, simultaneously,
and consider the
intercept term in the $T_L(q)$ fit
and the deviations of $T_L(q)$ ($L=L_{\min},...$) from the intercept,
in a $\chi^2$ test \cite{Herdan1955,9780898744149}.
The best fit is determined for $L_{\min}>4$ from which the $p$ value becomes monotonically
 increasing. The corresponding $L_{\min}$ is denoted by $L_{\min}^{\text {best}}$.
Since it is assumed [and evidently from Figs.\ \ref{fig:energyPDF_3models}b and \ref{fig:Tc} correct] that the exponents involved in the scaling
laws of $c_L^{\max}(q)$ and $T_L(q)$ are not independent, it is reasonable that
$L_{\min}^{\text{best}}$ simultaneously serves for the best fit of $c_L^{\max}(q)$. As observed in Fig.\ \ref{fig:energyPDF_3models}b, for $q=3$ it is plausible
 to try the ansatz $c_L^{\max} \approx (\ln L)^{\alpha/\nu}$ for the specific heat maximum. For the distance between $T_L$ and the infinite volume critical point, we use
$T_L - T_c \propto  L^{-1/\nu}(\ln L)^{\alpha/\nu}$ \cite{Kenna2006} and assume
 $\alpha,\nu$ satisfy the hyperscaling relation
\begin{equation}
\label{eq:hyper}
d\nu = 2-\alpha.
\end{equation}
The goodness-of-fit test yields $\chi^2/\text{d.o.f} = 1.14/7$, 
a $p$ value of 0.021 
and $L_{\min}^{\text{best}}= 20$ 
[from now on we will give for each $T_L(q)$ fit
its corresponding $\chi^2/\text{d.o.f}$, followed by the $p$ value and
$L_{\min}^{\text{best}}$, in parenthesis].
The intercept term in the $T_L(3)$ fit [Fig.\ \ref{fig:Tc}a] is $0.827(9)$ 
and $\alpha/\nu \approx 2.197(5)$. 
The $q=4$ model displays a pronounced power-law scaling. 
Assuming a second order scaling law 
$c_L^{\rm max} \propto L^{\alpha/\nu}(1 + AL^{-\omega}+o(L^{-\omega}))$, we focus on a correction to the leading order term.
The distance between $T_L(4)$ and $T_c$ scales (to leading order) as $L^{-1/\nu}$.
Again, next-to-leading-order unknown correction terms apparently involved.
A fit to a power-law decay of $L$ yields an intercept term $0.689(9)$ ($1.72/7,0.044,32$). 
The specific heat maximum scales as $L^{1.832(7)}$. The picture is different when $q=5$.
The rather asymptotic behaviour of
the energy PDF as shown in Fig.\ \ref{fig:energyPDF_3models}a
suggests the $q=5$ data are compatible with the first-order transition
volume dependent scaling laws.
The conventional $T_L-T_c\propto L^{-d}$ fit gives
$T_c(5)\approx 0.606(1)$ ($2.08/8,0.033,16$). 
A log-log fit to $c_L^{\max}$ against $L$,
for $L\geq 16$ gives a slope $1.992(6)$, so 
a volume-dependent scaling for the specific heat maximum is indeed conceivable.
To further support a second-order behaviour when $q=4$
we consider the universal scaling form
\begin{equation}
\label{eq:Cvscacling}
c_L = L^{\alpha/\nu}{\cal F}(tL^{1/\nu}),
\end{equation}
where ${\cal F}(x)$ is a universal scaling function of the dimensionless variable $x=tL^{1/\nu}$ and $t=(T-T_c)/T_c$ is the reduced temperature.
As clearly shown in Fig.\ \ref{fig:CvUniversal}, the specific heat, normalized by $L^{\alpha/\nu}$, collapses on a single curve as follows from (\ref{eq:Cvscacling}).
Thus, it is reasonable to assume the hyperscaling relation
indeed holds, in consistency with the scaling relations we use.

\begin{figure}
\includegraphics[width = \columnwidth,height = 6.7cm]{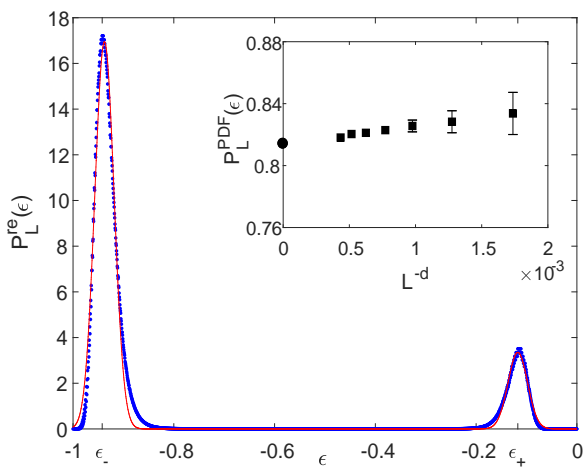}
\caption{\label{fig:energyPDF}
Reweighted PDF \cite{Janke1993} (blue symbols) together with a double Gaussian fit
for $L=44$. Note that the peaks are centred at points satisfying
$P_L(\epsilon_-)\approx qP_L(\epsilon_+)$.
The inset shows the difference between these points, as a function of $L^{-d}$ (closed squares).
Absent error bars are smaller than the symbols.
The estimated infinite volume 
$\Delta\epsilon_\infty^{\rm PDF}\approx 0.813(9)$ 
is denoted by the closed circle.
 Lattices with $L<24$ have too noisy distributions around the peaks and are therefore omitted.}
\end{figure}

Another manifestation of the $q=5$ discontinuous transition is the latent heat, estimated in two different ways.
First, by measuring the distance between the locations
of the peaks in a Gaussian fit to the energy PDF (Fig.\ \ref{fig:energyPDF})
and then trying the ansatz 
$\Delta \epsilon_L^{\rm PDF}\approx\Delta \epsilon^{\rm PDF}_\infty + \text{const}\times L^{-d}$, and second, using \cite{Challa1986}
\begin{equation}
c_L^{\max}\approx \frac{(\epsilon_+-\epsilon_-)^2}{4T_c^2}L^d+\frac{c_++c_-}{2},
\label{eq:Latent}
\end{equation}
where $c_+,c_-$ are temperature independent terms.
The PDF fit, for $L\geq 24$, produces
$\Delta \epsilon^{\rm PDF}_\infty =
\epsilon_+^{\rm PDF}-\epsilon_-^{\rm PDF} \approx 0.813(9)$ 
($\chi^2/\text{d.o.f}=1.35/6,p=0.058$) 
while\ (\ref{eq:Latent}), choosing
$T_c(5)\approx 0.606(1)$, yields $\Delta\epsilon =\epsilon_+-\epsilon_- \approx 0.809(5)$. 
The two results reasonably agree.

To conclude, we turn to test our analytical predictions against some of the simulations results.
First we compare the zeroth-order bounds with the simulations predictions. The results are summarized in Table \ref{tab:table1}.
As expected,\ (\ref{eq:Tc}) becomes a better approximation as $q$ grows.
Next, having in mind that for $q=5$ the transition is first order,
we give a lower bound on the correlation length $\xi(5)$
with the help of\ (\ref{eq:ratio}) and\ (\ref{eq:Tc_xi}). Taking
$\hat T_c\approx 0.606(1)$ 
we obtain $\xi(5) > 81$. 
This result justifies our FSS analysis in the sense that
the lattice sizes we use are compatible with $\xi(5)$.

\begin{table}[h]
\caption{\label{tab:table1}Estimates of the transition temperatures for the three models, using
the zeroth-order bound\ (\ref{eq:Tc}) and the simulations results. The relative error is
given in the last column. The supplementary $q=10$ result is
based on additional simulations for lattices with $4\leq L\leq 36$ and
a $T_L-T_c\propto L^{-d}$ fit ($2.75/7,0.084,8$).} 
\begin{ruledtabular}
\begin{tabular}{cccc}
$q$              & Bound & Simulations & Error (\%) \\
\hline
3 & 0.910(2) & 0.827(9) & 9.9 \\ 
4  &0.721(3) & 0.689(9) & 4.6   \\
5  & 0.621(3) & 0.606(1) & 2.5 \\
10 & 0.434(2) & 0.432(4) & 0.4\\ 
\end{tabular}
\end{ruledtabular}
\end{table}	

\section{conclusions}
\label{sec:con}
The transition nature of the FPS is controlled by
large scale lattice animals.
Based on the lattice animals asymptotic growth,
the transition is found continuous for $q\leq 4$ and discontinuous for $q>4$.
The $q=4$ is borderline. In the case in which the assumption that typical large clusters have
(to leading order) the same number of sites and faces breaks down (e.g.,
when the number of clusters satisfying $\lim_{k\to\infty}\frac{m_k}{k}>1$
is sufficiently large), the $q=4$ model might undergo a first-order transition. It is expected that large animals growth controls the transition order in other lattices as well. Specifically, it is known \cite{Barequet2017} that
the asymptotic number of triangular animals (polyamonds) of size $k$, $a_k$, satisfies $\lim_{k\to\infty}\sqrt[k]a_k = \lambda_t$
with $2.8424 < \lambda_t < 3.6050$. The number of faces in a typical large cluster
is (to leading order) twice the number of sites. Thus, the transition
is continuous $\it at~least$ for $q\leq 4$. Moreover, it can be easily shown the transition point is no larger than $2/\ln q$. The WL simulations and FSS analysis confirm our analytical predictions.
That is, the $q=3$ model displays a scaling behaviour typical to a second-order transition
and the $q=5$ numerical footprints are significantly first order.
While the $q=3$ FSS shows a very slow approach to the asymptotic regime, the $q=5$ sample sizes are compatible with $\xi(5)$.
The $\chi^2$ goodness-of-fit tests support the scaling laws we use. In particular,
for $q=3$ it follows that the free-energy is homogeneous in the small-$L$ regime, since
the critical indices apparently obey (up to small corrections)\ (\ref{eq:hyper}).
The $q=4$ model is rather unique. The double-peaked shape of the energy distribution is also observed in
models exhibiting a relatively weak first-order transition such as the $q=8$ usual Potts model,
[see Fig.\ 1c in \cite{Janke1993}]. On the other hand, Fig.\ \ref{fig:CvUniversal} remarkably
confirms\ (\ref{eq:Cvscacling}),
suggesting a divergence of the correlation length $\xi(4)\propto|t|^{-\nu}$ as $t\to 0$.
The indefiniteness of the four-state model manifested both analytically and numerically, is in agreement with
renormaliztion group (RG) predictions. The dynamics of models lying in the universality class of the two-site interaction $q=4$ Potts model (TSP)
flows towards the multicritical point $q_c=4$ \cite{Nienhuis1979,Nauenberg1980,Cardy1980}.
However, a certain choice of parameters \cite{Blte2017} may drive the dynamics in
some of these models away from $q_c$, to the first-order domain.
In other words, in the marginal $q=4$ case, the transition nature (first versus second order) is sensitive to the model's details \cite{Blte2017}.
The lattice animals mechanism suggests that FPS may belong to the TSP universality class.
Nevertheless, it leaves room for a first-order-like RG description.
It should be emphasized that unlike the RG method which 
makes assumptions on the model under scaling, our approach is direct and fundamental, building on first principles, and thus, we think, 
is preferable to RG for the studied question.
As a concluding remark, we believe that being general, our theoretical framework can be extended to other lattices, more complicated Hamiltonians and higher
dimensions.

\begin{acknowledgments}
We wish to thank Gidi Amir for fruitful discussions.
We also thank the anonymous referees reviewing the manuscript for useful comments and suggestions.
\end{acknowledgments}
\appendix
\section{The critical point}
\label{app:Tc}
\subsection{Derivation of equation\ (\ref{eq:Lambda})}
We give a detailed derivation of\ (\ref{eq:Lambda}) yielding the critical temperature\ (\ref{eq:Tc_mod}).
Since\ (\ref{eq:Tc_mod}) is also useful in estimating the finite correlation length
in the first-order case [see.\ (\ref{eq:xi}) and Appendix\ \ref{app:xi}], the derivation concerns with this class of models.
However, it is stressed that\ (\ref{eq:Tc_mod}) holds for arbitrary $q$.

Let $\epsilon_n$ be a sequence of positive small numbers. Then there exist
a sequence $k(\epsilon_n)$ and sets
\begin{equation}
\label{eq:kappa_n}
\kappa_n = \left\lbrace ~k > k(\epsilon_n)~: ~\left |\frac{\sum_{m_k}{\cal G}(k,m_k)}{c\lambda^k/k}-1\right | < \epsilon_n\right\rbrace,
\end{equation}
associated with animals ${\cal G}(k,m_k)$ with $k$ faces
and $m_k$ sites in the asymptotic regime.
Consider further, for every $n$, the set $A_n$ of {\it all} the animals
with an asymptotic $k$
\begin{equation}
\label{eq:setA}
A_n = \left\lbrace ~{\cal G}(k,m_k)~:~k\in \kappa_n\right\rbrace.
\end{equation}
We now define the (small) class of large-$k$ simple animals
\begin{equation}
\label{eq:classA}
\hat A = \left\lbrace {\cal G}(k,m_k)\in\bigcup_n A_n~:~\frac{m_k-k}{\sqrt k }\leq B\right\rbrace,
\end{equation}
where $B$ is a positive constant. Equations\ (\ref{eq:kappa_n})-(\ref{eq:classA}) allow us to define
\begin{equation}
\label{eq:sup}
\theta = \sup_k\left(\sup_{m_k:~{\cal G}(k,m_k)\in \hat A}\frac{m_k}{k}\right).
\end{equation}
Next, let
$r_j,~j=1,2,...,j_{\max}\leq {\cal N},~{\cal N}\in\bigcup_n\kappa_n$ be a sequence
satisfying  $\frac{1}{{\cal N}} < r_j<\frac{2}{{\cal N}}$. Construct another sequence with $j_{\max}$ integers $k_j\leq {\cal N}$ from
$\bigcup_n\kappa_n$. Define now for every $1\leq j\leq j_{\max}$
\begin{equation}
\hat{A_j}=\left\lbrace {\cal G}(k_j,m_{k_j})\in \hat A~ \text{s.t}~ \frac{m_{k_j}}{k_j} >
\theta-r_j\right\rbrace.
\end{equation}
Take $Z_{N}^{\rm low}\leq \hat Z_{{\cal N}}^{\rm low}$ where
\begin{eqnarray}
\label{eq:Z_low}
\hat Z_{{\cal N}}^{\rm low}\propto q^{{\cal N}}\sum_{j}\sum_{m_{k_j}}{\cal G}(k_j,m_{k_j})q^{-m_{k_j}}u^{k_j}
 \nonumber \\ \leq
q^{{\cal N}}\sum_{j}\sum_{m_{k_j}}{\cal G}(k_j,m_{k_j})\left(\frac{u}{q^{\theta-r_j}}\right)^{k_j} \nonumber \\ \leq
q^{{\cal N}}\sum_{j}\hat g_{k_j}\left(\frac{u}{q^{\theta-2/{{\cal N}}}}\right)^{k_j}
 \nonumber \\
\leq 
q^{{\cal N}}\left [K{\cal N} \binom{{\cal N}}{a\sqrt {{\cal N}}}\left(\frac{u}{q^{\theta-2/{{\cal N}}}}\right)^{{\cal N}}+o(\lambda^{\cal N})\right ].
\end{eqnarray}
The $m_{k_{j}}$ summations in\ (\ref{eq:Z_low}) taken over site variables of animals in $\hat{A_j}$, satisfy
\begin{equation}
\sum_{m_{k_j}}{\cal G}(k_j,m_{k_j})\leq \hat g_{k_j}.
\end{equation}
Since $\hat g_{k_j}$ count simple animals, their contributions to the 
leading order term are no larger than
$K\binom{{\cal N}}{a\sqrt {{\cal N}}}$ where $K,a$ are constants.
It follows immediately from\ (\ref{eq:Z_low}) that
\begin{eqnarray}
\label{eq:u_c}
\lim_{{\cal N}\to\infty}(\hat Z_{{\cal N}}^{\rm low})^{1/{{\cal N}}}
=\lim_{N\to\infty}(Z_{N}^{\rm low})^{1/N}=uq^{1-\theta}.
\end{eqnarray}

\subsection{Equation\ (\ref{eq:sup0}) and first-order transitions}
When the system undergoes a first-order phase transition,
$q$ ordered states coexist with a single disordered state at the critical point.
In\ (\ref{eq:sup0}) we utilize this as follows.
Consider a simple large animal with $k=\alpha N~(\alpha<1)$ faces and $m_k$ sites.
Then, the change in the free energy when making a macroscopic number of finite clusters monochromatic may be written
\begin{eqnarray}
\label{eq:DelF_high}
\Delta F(k,m_k,T) = N[-(1-\alpha)+\sigma(1-\alpha)\nonumber\\
+(1-\alpha\text{\footnotesize $\frac{m_k}{k}$})T\ln q]+ \text {h.o.t},
\end{eqnarray}
where $0<\sigma<1$ controls the energy loss due to boundary interactions of the finite clusters.
Applying now\ (\ref{eq:sup}) to (\ref{eq:DelF_high}) gives 
$\Delta F_u(T)\leq \Delta F(k,m_k,T) $ with 
\begin{eqnarray}
\label{eq:DelF_high_min}
\Delta F_u(T) = N[-(1-\alpha)(1-\sigma)\nonumber\\
+(1-\alpha\theta)T\ln q] + \text {h.o.t}.
\end{eqnarray}
Equation\ (\ref{eq:DelF_high_min}) holds provided the leading order term vanishes at the critical
point. In addition,\ (\ref{eq:DelF_high_min}) should be {\it unstable} in some left neighbourhood of $T_c$. These can be established first by
taking $\Delta F_u(T_c)=\text {h.o.t}$ for $T_c=\hat T_c = 1/(\theta\ln q)$, leading to
\begin{equation}
\label{eq:theta1}
\theta = \frac{1}{1-\sigma(1-\alpha)}.
\end{equation}
Second, consider $\Delta F_s(T)$, the
free energy change due to the formation of a single giant component, given by
\begin{equation}
\label{eq:Delta_F_s}
\Delta F_s(T) = N(-\alpha + \alpha\theta T\ln q) + \text {h.o.t}.
\end{equation}
Plugging (\ref{eq:theta1}) into (\ref{eq:DelF_high_min}) and (\ref{eq:Delta_F_s})
it follows that $\Delta F_s(T_c^-)< \Delta F_u(T_c^-)$ if and only if
\begin{equation}
\label{eq:treshold}
\alpha >\frac{1}{2\theta}.
\end{equation}
Equations\ (\ref{eq:DelF_high_min})-(\ref{eq:treshold}) assert that when a (first-order) phase transition occurs,
the fraction of faces constructing a monochromatic GC is
no smaller than $1/2\theta$. It should be noted that the critical threshold $\alpha_c = 1/2\theta$ increases with $q$ (see Appendix\ \ref{app:xi}) in accordance with the system's attempt to reduce entropy.

We conclude by stating that\ (\ref{eq:Lambda}) [and so\ (\ref{eq:Tc_mod})]
holds for the second-order models as well
In order that the number of animals with $k$ faces to be maximal,
the system picks those with a maximal number of sites. Equation\ (\ref{eq:sup0}) then immediately follows.
In addition, constructing
$\theta$, fractal animal are involved so that $\hat A$ in\ (\ref{eq:sup}) may be replaced with
${\cal \hat A}\subseteq\bigcup_n A_n$ \footnote{We take
${\cal \hat A}\subseteq\bigcup_n A_n$ to
make sure the inner supremum in\ (\ref{eq:sup}) exists}.
\section{The correlation length}
\label{app:xi}
In the following, we derive the relation between the first-order model finite correlation length and
the critical temperature, formulated by\ (\ref{eq:xi}).
Observe that for animals in $\hat A$,\ (\ref{eq:classA}) implies
\begin{equation}
\frac{m_k}{k}\leq 1+\frac{\hat c}{\sqrt{k}}+...
\end{equation}
Hence there exist a sequence $\hat k_n\leq k(\epsilon_n)$ such that
\begin{equation}
\theta \leq 1+\frac{\hat c}{\sqrt{\hat k_n}}+...,
\end{equation}
leading to
\begin{equation}
\label{eq:theta_xi_app}
\theta = 1+\frac{c_1}{\xi}+... = \inf_n\left(1+\frac{\hat c}{\sqrt{\hat k_n}}+...\right),
\end{equation}
with $[\xi^2] = \max_n(\hat k_n)$ and $c_1 = \hat c$.
The correlation length, as follows from\ (\ref{eq:theta_xi_app}), may be interpreted as
a typical length measuring large finite domains.
Writing the RHS of\ (\ref{eq:xi}) as a power series
$\sum_{n=0}^\infty c_n x^{n}$ at $x=\xi^{-1}$, it follows from\ (\ref{eq:theta1}) that
$\lim_{n\to\infty}\sqrt[n]{c_n} = \xi\sigma(1-\alpha)$ so the series indeed converges to $\theta$.

Observe that the above analysis can be extended to arbitrary $q$ first-order systems. We expect that as $q$ grows the deviations from a perfect square critical giant component
 become smaller. This may be formulated by constructing subclasses $\hat A(q)\subseteq \hat A$ with animals ${\cal G}(k,m_k)$ satisfying $\sup_k\frac{m_k-k}{\sqrt k}=B(q)$,
where the constants $B(q)$ are expected to decrease with $q$.
Replacing $\hat A$ in\ (\ref{eq:sup}) with $\hat A(q)$, $\theta$ essentially becomes $q$ dependent. It acquires lower values
as $q$ grows, as also realized in Table \ref{tab:table1}, where the simulated temperature
approaches better the bound $1/\ln q$, when $q$ changes from $q=5$ to $q=10$.


 
%
 

\begin{thebibliography}{35}%
\makeatletter
\providecommand \@ifxundefined [1]{%
 \@ifx{#1\undefined}
}%
\providecommand \@ifnum [1]{%
 \ifnum #1\expandafter \@firstoftwo
 \else \expandafter \@secondoftwo
 \fi
}%
\providecommand \@ifx [1]{%
 \ifx #1\expandafter \@firstoftwo
 \else \expandafter \@secondoftwo
 \fi
}%
\providecommand \natexlab [1]{#1}%
\providecommand \enquote  [1]{``#1''}%
\providecommand \bibnamefont  [1]{#1}%
\providecommand \bibfnamefont [1]{#1}%
\providecommand \citenamefont [1]{#1}%
\providecommand \href@noop [0]{\@secondoftwo}%
\providecommand \href [0]{\begingroup \@sanitize@url \@href}%
\providecommand \@href[1]{\@@startlink{#1}\@@href}%
\providecommand \@@href[1]{\endgroup#1\@@endlink}%
\providecommand \@sanitize@url [0]{\catcode `\\12\catcode `\$12\catcode
  `\&12\catcode `\#12\catcode `\^12\catcode `\_12\catcode `\%12\relax}%
\providecommand \@@startlink[1]{}%
\providecommand \@@endlink[0]{}%
\providecommand \url  [0]{\begingroup\@sanitize@url \@url }%
\providecommand \@url [1]{\endgroup\@href {#1}{\urlprefix }}%
\providecommand \urlprefix  [0]{URL }%
\providecommand \Eprint [0]{\href }%
\providecommand \doibase [0]{http://dx.doi.org/}%
\providecommand \selectlanguage [0]{\@gobble}%
\providecommand \bibinfo  [0]{\@secondoftwo}%
\providecommand \bibfield  [0]{\@secondoftwo}%
\providecommand \translation [1]{[#1]}%
\providecommand \BibitemOpen [0]{}%
\providecommand \bibitemStop [0]{}%
\providecommand \bibitemNoStop [0]{.\EOS\space}%
\providecommand \EOS [0]{\spacefactor3000\relax}%
\providecommand \BibitemShut  [1]{\csname bibitem#1\endcsname}%
\let\auto@bib@innerbib\@empty
\bibitem [{\citenamefont {Potts}\ and\ \citenamefont {Domb}(1952)}]{Potts1952}%
  \BibitemOpen
  \bibfield  {author} {\bibinfo {author} {\bibfnamefont {R.~B.}\ \bibnamefont
  {Potts}}\ and\ \bibinfo {author} {\bibfnamefont {C.}~\bibnamefont {Domb}},\
  }\href {\doibase 10.1017/s0305004100027419} {\bibfield  {journal} {\bibinfo
  {journal} {Mathematical Proceedings of the Cambridge Philosophical Society}\
  }\textbf {\bibinfo {volume} {48}},\ \bibinfo {pages} {106} (\bibinfo {year}
  {1952})}\BibitemShut {NoStop}%
\bibitem [{\citenamefont {Wu}(1982)}]{Wu1982}%
  \BibitemOpen
  \bibfield  {author} {\bibinfo {author} {\bibfnamefont {F.~Y.}\ \bibnamefont
  {Wu}},\ }\href {\doibase 10.1103/revmodphys.54.235} {\bibfield  {journal}
  {\bibinfo  {journal} {Reviews of Modern Physics}\ }\textbf {\bibinfo {volume}
  {54}},\ \bibinfo {pages} {235} (\bibinfo {year} {1982})}\BibitemShut
  {NoStop}%
\bibitem [{\citenamefont {Baxter}\ \emph {et~al.}(1978)\citenamefont {Baxter},
  \citenamefont {Temperley},\ and\ \citenamefont {Ashley}}]{Baxter1978}%
  \BibitemOpen
  \bibfield  {author} {\bibinfo {author} {\bibfnamefont {R.~J.}\ \bibnamefont
  {Baxter}}, \bibinfo {author} {\bibfnamefont {H.~N.~V.}\ \bibnamefont
  {Temperley}}, \ and\ \bibinfo {author} {\bibfnamefont {S.~E.}\ \bibnamefont
  {Ashley}},\ }\href {\doibase 10.1098/rspa.1978.0026} {\bibfield  {journal}
  {\bibinfo  {journal} {Proceedings of the Royal Society A: Mathematical,
  Physical and Engineering Sciences}\ }\textbf {\bibinfo {volume} {358}},\
  \bibinfo {pages} {535} (\bibinfo {year} {1978})}\BibitemShut {NoStop}%
\bibitem [{\citenamefont {Enting}\ and\ \citenamefont {Wu}(1982)}]{Enting1982}%
  \BibitemOpen
  \bibfield  {author} {\bibinfo {author} {\bibfnamefont {I.~G.}\ \bibnamefont
  {Enting}}\ and\ \bibinfo {author} {\bibfnamefont {F.~Y.}\ \bibnamefont
  {Wu}},\ }\href {\doibase 10.1007/bf01012610} {\bibfield  {journal} {\bibinfo
  {journal} {Journal of Statistical Physics}\ }\textbf {\bibinfo {volume}
  {28}},\ \bibinfo {pages} {351} (\bibinfo {year} {1982})}\BibitemShut
  {NoStop}%
\bibitem [{\citenamefont {Wu}(1979)}]{Wu1979}%
  \BibitemOpen
  \bibfield  {author} {\bibinfo {author} {\bibfnamefont {F.~Y.}\ \bibnamefont
  {Wu}},\ }\href {\doibase 10.1088/0022-3719/12/17/002} {\bibfield  {journal}
  {\bibinfo  {journal} {Journal of Physics C: Solid State Physics}\ }\textbf
  {\bibinfo {volume} {12}},\ \bibinfo {pages} {L645} (\bibinfo {year}
  {1979})}\BibitemShut {NoStop}%
\bibitem [{\citenamefont {Wu}\ and\ \citenamefont {Lin}(1980)}]{Wu1980}%
  \BibitemOpen
  \bibfield  {author} {\bibinfo {author} {\bibfnamefont {F.~Y.}\ \bibnamefont
  {Wu}}\ and\ \bibinfo {author} {\bibfnamefont {K.~Y.}\ \bibnamefont {Lin}},\
  }\href {\doibase 10.1088/0305-4470/13/2/026} {\bibfield  {journal} {\bibinfo
  {journal} {Journal of Physics A: Mathematical and General}\ }\textbf
  {\bibinfo {volume} {13}},\ \bibinfo {pages} {629} (\bibinfo {year}
  {1980})}\BibitemShut {NoStop}%
\bibitem [{\citenamefont {Wu}\ and\ \citenamefont {Zia}(1981)}]{Wu1981}%
  \BibitemOpen
  \bibfield  {author} {\bibinfo {author} {\bibfnamefont {F.~Y.}\ \bibnamefont
  {Wu}}\ and\ \bibinfo {author} {\bibfnamefont {R.~K.~P.}\ \bibnamefont
  {Zia}},\ }\href {\doibase 10.1088/0305-4470/14/3/018} {\bibfield  {journal}
  {\bibinfo  {journal} {Journal of Physics A: Mathematical and General}\
  }\textbf {\bibinfo {volume} {14}},\ \bibinfo {pages} {721} (\bibinfo {year}
  {1981})}\BibitemShut {NoStop}%
\bibitem [{\citenamefont {Giri}\ \emph {et~al.}(1977)\citenamefont {Giri},
  \citenamefont {Stephen},\ and\ \citenamefont {Grest}}]{Giri1977}%
  \BibitemOpen
  \bibfield  {author} {\bibinfo {author} {\bibfnamefont {M.~R.}\ \bibnamefont
  {Giri}}, \bibinfo {author} {\bibfnamefont {M.~J.}\ \bibnamefont {Stephen}}, \
  and\ \bibinfo {author} {\bibfnamefont {G.~S.}\ \bibnamefont {Grest}},\ }\href
  {\doibase 10.1103/physrevb.16.4971} {\bibfield  {journal} {\bibinfo
  {journal} {Physical Review B}\ }\textbf {\bibinfo {volume} {16}},\ \bibinfo
  {pages} {4971} (\bibinfo {year} {1977})}\BibitemShut {NoStop}%
\bibitem [{\citenamefont {Kunz}\ and\ \citenamefont {Wu}(1978)}]{Kunz1978}%
  \BibitemOpen
  \bibfield  {author} {\bibinfo {author} {\bibfnamefont {H.}~\bibnamefont
  {Kunz}}\ and\ \bibinfo {author} {\bibfnamefont {F.~Y.}\ \bibnamefont {Wu}},\
  }\href {\doibase 10.1088/0022-3719/11/8/512} {\bibfield  {journal} {\bibinfo
  {journal} {Journal of Physics C: Solid State Physics}\ }\textbf {\bibinfo
  {volume} {11}},\ \bibinfo {pages} {L357} (\bibinfo {year}
  {1978})}\BibitemShut {NoStop}%
\bibitem [{\citenamefont {Burkhardt}(1979)}]{Burkhardt1979}%
  \BibitemOpen
  \bibfield  {author} {\bibinfo {author} {\bibfnamefont {T.~W.}\ \bibnamefont
  {Burkhardt}},\ }\href {\doibase 10.1103/physrevb.20.2905} {\bibfield
  {journal} {\bibinfo  {journal} {Physical Review B}\ }\textbf {\bibinfo
  {volume} {20}},\ \bibinfo {pages} {2905} (\bibinfo {year}
  {1979})}\BibitemShut {NoStop}%
\bibitem [{\citenamefont {Essam}(1979)}]{Essam1979}%
  \BibitemOpen
  \bibfield  {author} {\bibinfo {author} {\bibfnamefont {J.~W.}\ \bibnamefont
  {Essam}},\ }\href {\doibase 10.1063/1.524264} {\bibfield  {journal} {\bibinfo
   {journal} {Journal of Mathematical Physics}\ }\textbf {\bibinfo {volume}
  {20}},\ \bibinfo {pages} {1769} (\bibinfo {year} {1979})}\BibitemShut
  {NoStop}%
\bibitem [{\citenamefont {Baxter}(1973{\natexlab{a}})}]{Baxter1973a}%
  \BibitemOpen
  \bibfield  {author} {\bibinfo {author} {\bibfnamefont {R.~J.}\ \bibnamefont
  {Baxter}},\ }\href {\doibase 10.1088/0022-3719/6/23/005} {\bibfield
  {journal} {\bibinfo  {journal} {Journal of Physics C: Solid State Physics}\
  }\textbf {\bibinfo {volume} {6}},\ \bibinfo {pages} {L445} (\bibinfo {year}
  {1973}{\natexlab{a}})}\BibitemShut {NoStop}%
\bibitem [{\citenamefont {Baxter}(1973{\natexlab{b}})}]{Baxter1973b}%
  \BibitemOpen
  \bibfield  {author} {\bibinfo {author} {\bibfnamefont {R.~J.}\ \bibnamefont
  {Baxter}},\ }\href {\doibase 10.1007/bf01016845} {\bibfield  {journal}
  {\bibinfo  {journal} {Journal of Statistical Physics}\ }\textbf {\bibinfo
  {volume} {9}},\ \bibinfo {pages} {145} (\bibinfo {year}
  {1973}{\natexlab{b}})}\BibitemShut {NoStop}%
\bibitem [{\citenamefont {Wang}\ and\ \citenamefont
  {Landau}(2001{\natexlab{a}})}]{Wang2001}%
  \BibitemOpen
  \bibfield  {author} {\bibinfo {author} {\bibfnamefont {F.}~\bibnamefont
  {Wang}}\ and\ \bibinfo {author} {\bibfnamefont {D.~P.}\ \bibnamefont
  {Landau}},\ }\href {\doibase 10.1103/physrevlett.86.2050} {\bibfield
  {journal} {\bibinfo  {journal} {Physical Review Letters}\ }\textbf {\bibinfo
  {volume} {86}},\ \bibinfo {pages} {2050} (\bibinfo {year}
  {2001}{\natexlab{a}})}\BibitemShut {NoStop}%
\bibitem [{\citenamefont {Wang}\ and\ \citenamefont
  {Landau}(2001{\natexlab{b}})}]{Wang2001a}%
  \BibitemOpen
  \bibfield  {author} {\bibinfo {author} {\bibfnamefont {F.}~\bibnamefont
  {Wang}}\ and\ \bibinfo {author} {\bibfnamefont {D.~P.}\ \bibnamefont
  {Landau}},\ }\href {\doibase 10.1103/physreve.64.056101} {\bibfield
  {journal} {\bibinfo  {journal} {Physical Review E}\ }\textbf {\bibinfo
  {volume} {64}},\ \bibinfo {pages} {6101} (\bibinfo {year}
  {2001}{\natexlab{b}})}\BibitemShut {NoStop}%
\bibitem [{\citenamefont {Martin-Mayor}(2007)}]{MartinMayor2007}%
  \BibitemOpen
  \bibfield  {author} {\bibinfo {author} {\bibfnamefont {V.}~\bibnamefont
  {Martin-Mayor}},\ }\href {\doibase 10.1103/physrevlett.98.137207} {\bibfield
  {journal} {\bibinfo  {journal} {Physical Review Letters}\ }\textbf {\bibinfo
  {volume} {98}},\ \bibinfo {pages} {7207} (\bibinfo {year}
  {2007})}\BibitemShut {NoStop}%
\bibitem [{\citenamefont {Baxter}(2013)}]{0486462714}%
  \BibitemOpen
  \bibfield  {author} {\bibinfo {author} {\bibfnamefont {R.~J.}\ \bibnamefont
  {Baxter}},\ }\href@noop {} {\emph {\bibinfo {title} {Exactly Solved Models in
  Statistical Mechanics (Dover Books on Physics)}}}\ (\bibinfo  {publisher}
  {Dover Publications},\ \bibinfo {year} {2013})\BibitemShut {NoStop}%
\bibitem [{\citenamefont {Barequet}\ \emph {et~al.}(2016)\citenamefont
  {Barequet}, \citenamefont {Rote},\ and\ \citenamefont
  {Shalah}}]{Barequet2016}%
  \BibitemOpen
  \bibfield  {author} {\bibinfo {author} {\bibfnamefont {G.}~\bibnamefont
  {Barequet}}, \bibinfo {author} {\bibfnamefont {G.}~\bibnamefont {Rote}}, \
  and\ \bibinfo {author} {\bibfnamefont {M.}~\bibnamefont {Shalah}},\ }\href
  {\doibase 10.1145/2851485} {\bibfield  {journal} {\bibinfo  {journal}
  {Communications of the {ACM}}\ }\textbf {\bibinfo {volume} {59}},\ \bibinfo
  {pages} {88} (\bibinfo {year} {2016})}\BibitemShut {NoStop}%
\bibitem [{\citenamefont {Klarner}\ and\ \citenamefont
  {Rivest}(1973)}]{Klarner1973}%
  \BibitemOpen
  \bibfield  {author} {\bibinfo {author} {\bibfnamefont {D.~A.}\ \bibnamefont
  {Klarner}}\ and\ \bibinfo {author} {\bibfnamefont {R.~L.}\ \bibnamefont
  {Rivest}},\ }\href {\doibase 10.4153/cjm-1973-060-4} {\bibfield  {journal}
  {\bibinfo  {journal} {Journal canadien de math{\'{e}}matiques}\ }\textbf
  {\bibinfo {volume} {25}},\ \bibinfo {pages} {585} (\bibinfo {year}
  {1973})}\BibitemShut {NoStop}%
\bibitem [{\citenamefont {Jensen}(2003)}]{Jensen2003}%
  \BibitemOpen
  \bibfield  {author} {\bibinfo {author} {\bibfnamefont {I.}~\bibnamefont
  {Jensen}},\ }in\ \href {\doibase 10.1007/3-540-44863-2_21} {\emph {\bibinfo
  {booktitle} {Lecture Notes in Computer Science}}}\ (\bibinfo  {publisher}
  {Springer Berlin Heidelberg},\ \bibinfo {year} {2003})\ pp.\ \bibinfo {pages}
  {203--212}\BibitemShut {NoStop}%
\bibitem [{Note1()}]{Note1}%
  \BibitemOpen
  \bibinfo {note} {It can be shown that $k$ clusters and clusters of size
  $o(N)$ do not share mutual $\protect \text {"corner"}$ sites, asymptotically
  almost surely.}\BibitemShut {Stop}%
\bibitem [{\citenamefont {Delfino}\ and\ \citenamefont
  {Tartaglia}(2017)}]{Delfino2017}%
  \BibitemOpen
  \bibfield  {author} {\bibinfo {author} {\bibfnamefont {G.}~\bibnamefont
  {Delfino}}\ and\ \bibinfo {author} {\bibfnamefont {E.}~\bibnamefont
  {Tartaglia}},\ }\href@noop {} {\bibfield  {journal} {\bibinfo  {journal}
  {Physical Review E}\ }\textbf {\bibinfo {volume} {96}},\ \bibinfo {pages}
  {2137} (\bibinfo {year} {2017})}\BibitemShut {NoStop}%
\bibitem [{\citenamefont {Newman}\ and\ \citenamefont
  {Barkema}(1999)}]{Newman1999}%
  \BibitemOpen
  \bibfield  {author} {\bibinfo {author} {\bibfnamefont {M.~E.~J.}\
  \bibnamefont {Newman}}\ and\ \bibinfo {author} {\bibfnamefont {G.~T.}\
  \bibnamefont {Barkema}},\ }\href@noop {} {\emph {\bibinfo {title} {Monte
  Carlo Methods in Statistical Physics}}}\ (\bibinfo  {publisher} {Clarendon
  Press},\ \bibinfo {year} {1999})\BibitemShut {NoStop}%
\bibitem [{\citenamefont {Stanley}(1987)}]{Stanley1987}%
  \BibitemOpen
  \bibfield  {author} {\bibinfo {author} {\bibfnamefont {H.~E.}\ \bibnamefont
  {Stanley}},\ }\href@noop {} {\emph {\bibinfo {title} {Introduction to Phase
  Transitions and Critical Phenomena (International Series of Monographs on
  Physics)}}}\ (\bibinfo  {publisher} {Oxford University Press},\ \bibinfo
  {year} {1987})\BibitemShut {NoStop}%
\bibitem [{\citenamefont {Janke}(1993)}]{Janke1993}%
  \BibitemOpen
  \bibfield  {author} {\bibinfo {author} {\bibfnamefont {W.}~\bibnamefont
  {Janke}},\ }\href {\doibase 10.1103/physrevb.47.14757} {\bibfield  {journal}
  {\bibinfo  {journal} {Physical Review B}\ }\textbf {\bibinfo {volume} {47}},\
  \bibinfo {pages} {14757} (\bibinfo {year} {1993})}\BibitemShut {NoStop}%
\bibitem [{\citenamefont {Herdan}(1955)}]{Herdan1955}%
  \BibitemOpen
  \bibfield  {author} {\bibinfo {author} {\bibfnamefont {G.}~\bibnamefont
  {Herdan}},\ }\href {\doibase 10.1002/pol.1955.120178417} {\bibfield
  {journal} {\bibinfo  {journal} {Journal of Polymer Science}\ }\textbf
  {\bibinfo {volume} {17}},\ \bibinfo {pages} {315} (\bibinfo {year}
  {1955})}\BibitemShut {NoStop}%
\bibitem [{\citenamefont {Evans}(1982)}]{9780898744149}%
  \BibitemOpen
  \bibfield  {author} {\bibinfo {author} {\bibfnamefont {R.~D.}\ \bibnamefont
  {Evans}},\ }\href@noop {} {\emph {\bibinfo {title} {Atomic Nucleus}}}\
  (\bibinfo  {publisher} {Krieger Pub Co},\ \bibinfo {year} {1982})\BibitemShut
  {NoStop}%
\bibitem [{\citenamefont {Kenna}\ \emph {et~al.}(2006)\citenamefont {Kenna},
  \citenamefont {Johnston},\ and\ \citenamefont {Janke}}]{Kenna2006}%
  \BibitemOpen
  \bibfield  {author} {\bibinfo {author} {\bibfnamefont {R.}~\bibnamefont
  {Kenna}}, \bibinfo {author} {\bibfnamefont {D.~A.}\ \bibnamefont {Johnston}},
  \ and\ \bibinfo {author} {\bibfnamefont {W.}~\bibnamefont {Janke}},\
  }\href@noop {} {\bibfield  {journal} {\bibinfo  {journal} {Physical Review
  Letters}\ }\textbf {\bibinfo {volume} {96}},\ \bibinfo {pages} {5701}
  (\bibinfo {year} {2006})}\BibitemShut {NoStop}%
\bibitem [{\citenamefont {Challa}\ \emph {et~al.}(1986)\citenamefont {Challa},
  \citenamefont {Landau},\ and\ \citenamefont {Binder}}]{Challa1986}%
  \BibitemOpen
  \bibfield  {author} {\bibinfo {author} {\bibfnamefont {M.~S.~S.}\
  \bibnamefont {Challa}}, \bibinfo {author} {\bibfnamefont {D.~P.}\
  \bibnamefont {Landau}}, \ and\ \bibinfo {author} {\bibfnamefont
  {K.}~\bibnamefont {Binder}},\ }\href {\doibase 10.1103/physrevb.34.1841}
  {\bibfield  {journal} {\bibinfo  {journal} {Physical Review B}\ }\textbf
  {\bibinfo {volume} {34}},\ \bibinfo {pages} {1841} (\bibinfo {year}
  {1986})}\BibitemShut {NoStop}%
\bibitem [{\citenamefont {Barequet}\ \emph {et~al.}(2017)\citenamefont
  {Barequet}, \citenamefont {Shalah},\ and\ \citenamefont
  {Zheng}}]{Barequet2017}%
  \BibitemOpen
  \bibfield  {author} {\bibinfo {author} {\bibfnamefont {G.}~\bibnamefont
  {Barequet}}, \bibinfo {author} {\bibfnamefont {M.}~\bibnamefont {Shalah}}, \
  and\ \bibinfo {author} {\bibfnamefont {Y.}~\bibnamefont {Zheng}},\ }in\ \href
  {\doibase 10.1007/978-3-319-62389-4_5} {\emph {\bibinfo {booktitle} {Lecture
  Notes in Computer Science}}}\ (\bibinfo  {publisher} {Springer International
  Publishing},\ \bibinfo {year} {2017})\ pp.\ \bibinfo {pages}
  {50--61}\BibitemShut {NoStop}%
\bibitem [{\citenamefont {Nienhuis}\ \emph {et~al.}(1979)\citenamefont
  {Nienhuis}, \citenamefont {Berker}, \citenamefont {Riedel},\ and\
  \citenamefont {Schick}}]{Nienhuis1979}%
  \BibitemOpen
  \bibfield  {author} {\bibinfo {author} {\bibfnamefont {B.}~\bibnamefont
  {Nienhuis}}, \bibinfo {author} {\bibfnamefont {A.~N.}\ \bibnamefont
  {Berker}}, \bibinfo {author} {\bibfnamefont {E.~K.}\ \bibnamefont {Riedel}},
  \ and\ \bibinfo {author} {\bibfnamefont {M.}~\bibnamefont {Schick}},\ }\href
  {\doibase 10.1103/physrevlett.43.737} {\bibfield  {journal} {\bibinfo
  {journal} {Physical Review Letters}\ }\textbf {\bibinfo {volume} {43}},\
  \bibinfo {pages} {737} (\bibinfo {year} {1979})}\BibitemShut {NoStop}%
\bibitem [{\citenamefont {Nauenberg}\ and\ \citenamefont
  {Scalapino}(1980)}]{Nauenberg1980}%
  \BibitemOpen
  \bibfield  {author} {\bibinfo {author} {\bibfnamefont {M.}~\bibnamefont
  {Nauenberg}}\ and\ \bibinfo {author} {\bibfnamefont {D.~J.}\ \bibnamefont
  {Scalapino}},\ }\href {\doibase 10.1103/physrevlett.44.837} {\bibfield
  {journal} {\bibinfo  {journal} {Physical Review Letters}\ }\textbf {\bibinfo
  {volume} {44}},\ \bibinfo {pages} {837} (\bibinfo {year} {1980})}\BibitemShut
  {NoStop}%
\bibitem [{\citenamefont {Cardy}\ \emph {et~al.}(1980)\citenamefont {Cardy},
  \citenamefont {Nauenberg},\ and\ \citenamefont {Scalapino}}]{Cardy1980}%
  \BibitemOpen
  \bibfield  {author} {\bibinfo {author} {\bibfnamefont {J.~L.}\ \bibnamefont
  {Cardy}}, \bibinfo {author} {\bibfnamefont {M.}~\bibnamefont {Nauenberg}}, \
  and\ \bibinfo {author} {\bibfnamefont {D.~J.}\ \bibnamefont {Scalapino}},\
  }\href {\doibase 10.1103/physrevb.22.2560} {\bibfield  {journal} {\bibinfo
  {journal} {Physical Review B}\ }\textbf {\bibinfo {volume} {22}},\ \bibinfo
  {pages} {2560} (\bibinfo {year} {1980})}\BibitemShut {NoStop}%
\bibitem [{\citenamefont {Bl\"{o}te}\ \emph {et~al.}(2017)\citenamefont
  {Bl\"{o}te}, \citenamefont {Guo},\ and\ \citenamefont
  {Nightingale}}]{Blte2017}%
  \BibitemOpen
  \bibfield  {author} {\bibinfo {author} {\bibfnamefont {H.~W.~J.}\
  \bibnamefont {Bl\"{o}te}}, \bibinfo {author} {\bibfnamefont {W.}~\bibnamefont
  {Guo}}, \ and\ \bibinfo {author} {\bibfnamefont {M.~P.}\ \bibnamefont
  {Nightingale}},\ }\href {\doibase 10.1088/1751-8121/aa7b53} {\bibfield
  {journal} {\bibinfo  {journal} {Journal of Physics A: Mathematical and
  Theoretical}\ }\textbf {\bibinfo {volume} {50}},\ \bibinfo {pages} {4001}
  (\bibinfo {year} {2017})}\BibitemShut {NoStop}%
\bibitem [{Note2()}]{Note2}%
  \BibitemOpen
  \bibinfo {note} {We take ${\protect \cal \protect \mathaccentV
  {hat}05EA}\subseteq \DOTSB \bigcup@ \slimits@ _n A_n$ to make sure the inner
  supremum in\ (\ref {eq:sup}) exists}\BibitemShut {NoStop}%
\end{thebibliography}
\end{document}